# Domain wall nucleation in ferromagnetic nanowire with perpendicular magnetization stimulated by stray field of V-shaped magnetic particle

O. L. Ermolaeva, V. L. Mironov

Department of Magnetic Nanostructures,
Institute for Physics of Microstructures RAS, GSP-105, Nizhny Novgorod, 603950, Russia

We report the results of micromagnetic simulations of domain wall (DW) nucleation and pinning/depinning processes in ferromagnetic planar structure consisting of nanowire (NW) with perpendicular anisotropy and special V-shaped nanoparticle (NP) with in-plane anisotropy located on top of NW. The magnetization reversal features of this system in an external magnetic field are investigated depending on the direction of particle magnetic moment. Possible variants of magnetic logic cells (LCs) based on such system are discussed.

*Magnetic domain wall, ferromagnetic nanowire, domain wall nucleation, perpendicular anisotropy, magnetic logic*

## I. INTRODUCTION

Field–driven DW nucleation and controlled pinning/depinning processes in ferromagnetic NWs are the fundamental basis for the development of magnetic NW logic cells [1-13]. The information (logical "0" and "1") in these systems is encoded by magnetization direction in specified NW parts and logic computation is based on DWs motion in external magnetic field and their pinning/depinning at controlled gates. The promising variant of magnetic gates was discussed recently in Refs. [14-20] for planar NW systems with in-plane anisotropy. The main idea is the use of configurable stray fields generated by sets of bistable NPs placed near the NW to create controllable traps with variable DW pinning potential. These systems enable the realization of some LCs with different logical function [16-20]. An attempt to transfer this ideology at the planar system with perpendicular anisotropy has been made recently in Ref. [21]. Authors discussed the DW pinning in Co/Ni multilayer NW with perpendicular magnetization by magnetic stray field of permalloy rectangular NP (with in-plane magnetization) situated on top of NW. It was shown that the type of potential relief for DW motion and strength of DW pinning in NP stray field strongly depends on the orientation of NP magnetic moment relatively the direction of magnetization in adjacent NW domains. In the present paper we consider the peculiarities of DW nucleation and pinning in other system consisting of NW (with out of plane magnetization) and specially V-shaped ferromagnetic NP (with in-plane magnetization) located on top of NW. Possible application of this system for the realization of various LCs is discussed.

## II. NW-NP SYSTEM WITH RECTANGULAR NANOPARTICLE

As a model NW-PN system we investigated planar structure consisting of multilayer Co/Pt NW with perpendicular anisotropy and Co NP with in-plane anisotropy. The micromagnetic simulations were performed using Object Oriented MicroMagnetic Framework (OOMMF) software [22]. In order to understand the main peculiarities of magnetization reversal processes we considered firstly simple NW-NP system with rectangular Co NP on the top of Co/Pt NW (Fig. 1(a)). This model system had the following geometric parameters: Co/Pt NW was $1000 \times 100 \times 10$ nm$^3$, Co NP was $200 \times 100 \times 10$ nm$^3$. Magnetization in saturation was $M_{Co/Pt} = 0.8$ A/m and $M_{Co} = 1.4$ A/m; parameter of anisotropy was $K_{Co/Pt} = 6 \times 10^5$ J/m$^3$ and $K_{Co} = 0$; exchange stiffness was $A_{Co/Pt} = 1.5 \times 10^{-11}$ J/m and $A_{Co} = 2.5 \times 10^{-11}$ J/m [23]. In calculations the sample was divided into elementary cells of $5 \times 5 \times 1$ nm$^3$.

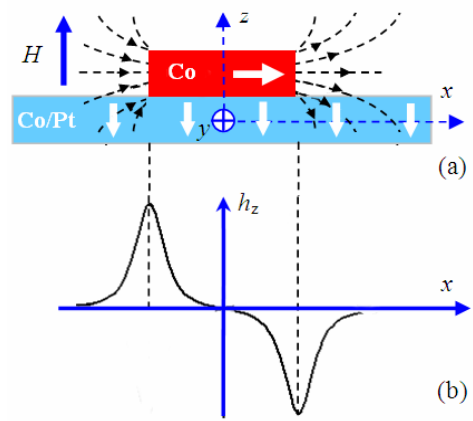

FIG. 1. (a) Schematic drawing of model NW-NP system consisting of rectangular Co NP with in-plane magnetization and Co/Pt NW with perpendicular magnetization. The magnetization in NP and NW is indicated by white arrows (b) The normed distribution of rectangular NP stray field ($h_z$) along the NW $x$ axis.

The energy of NW-NP interaction consists of exchange and magnetostatic terms. The schematic drawing of NP magnetic stray field distribution (z-component, $h_z$) is presented in Fig. 1(b). Since z-component of NP magnetic stray field coincides with direction of NW magnetization $\vec{M}$ at the NP right edge and has opposite direction at the left edge, the magnetostatic interaction increases the magnetization reversal barrier near NP right side and decreases at the left side. On the other hand, it is known that the exchange interaction between Co/Pt multilayer structure and covering Co film leads to the formation of noncollinear magnetization [24]. It can be used to control the dynamics of remagnetization in NP-NW system. The micromagnetic simulations shown that the exchange interaction between Co NP and Co/Pt NW leads to the distortion of NW and NP magnetization (see Fig. 2(b)). In particular, in case of strong exchange coupling the fractional remagnetization of NW is observed even without external field (Fig. 2(c)).

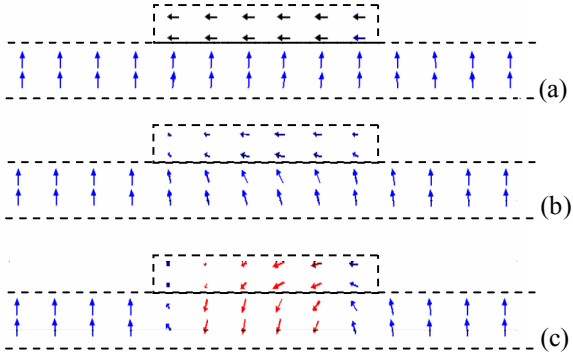

**FIG. 2.** Equilibrium distributions of magnetization in NW-NP system for different values of exchange stiffness $A_{Co/Pt-Co}$. (a) $A_{Co/Pt-Co} = 0$; (b) $A_{Co/Pt-Co} = 0.5 \cdot 10^{-11}$ J/m; (c) $A_{Co/Pt-Co} = 10^{-11}$ J/m.

As an example we simulated the dependence of z-component of NW magnetization $m_z$ (normed on $M_{Co/Pt}$) on external perpendicular magnetic field $H$ for exchange stiffness $A_{Co/Pt-Co} = 2 \times 10^{-11}$ J/m ($m_z(H)$ curve in Fig. 3(a)). An external magnetic field was directed upward and increased gradually from zero to 300 Oe with 10 Oe increment.

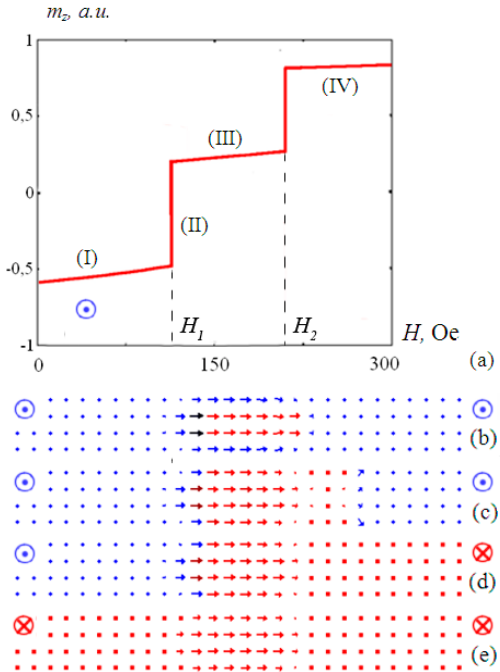

**FIG. 3.** (a) The dependence of normalized NW magnetization $m_z$ on external magnetic field in case of magnetostatic NP-NW interaction. (b)-(d) The distributions of NW magnetization corresponding to the sequence stages (I)-(IV) of magnetization reversal.

The corresponding sequential steps of magnetization reversal are demonstrated in Fig. 3(b-e). The color and special arrows indicate the direction of magnetization in NW. The blue dots (and circle with point) correspond to the upward, while the red dots (and circle with cross) correspond to the downward magnetization. The NW magnetization reversal occurs in two stages. At first step when the field reaches $H_1 = 110$ Oe the magnetization is changed in the area near NP right edge (Fig. 3(c)) and then the right part of NW is remagnetized (Fig. 3(d)). Afterward, when the field $H$ exceeds $H_2 = 210 > H_1$ the NW is remagnetized completely (Fig. 3(e)).

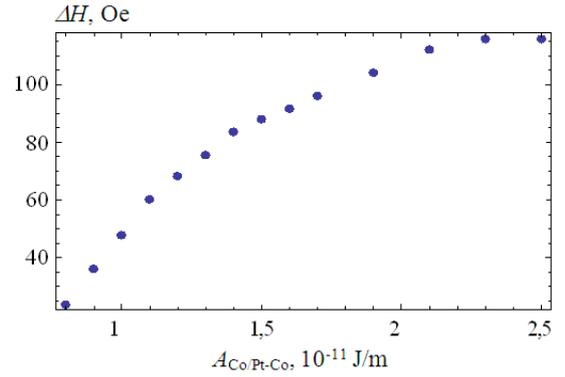

**FIG. 4.** The dependence of $\Delta H = H_2 - H_1$ on the Co/Pt-Co exchange stiffness.

The value of the difference between critical fields $\Delta H = H_2 - H_1$ is an important parameter defining the magnetization switching and LC operation. The dependence of $\Delta H$ on the exchange stiffness $A_{CoPt-Co}$ is presented in the Fig. 4. It is seen, that $\Delta H$ is increased with increasing of exchange coupling.

### III. NW-NP SYSTEM WITH V-SHAPED NANOPARTICLE

The shape of the NP plays an important role in the processes of magnetization reversal. To control the NW remagnetization we considered a special V-shaped NP with 90° corner (Fig. 5). The NW size was the same while NP area was $100 \times 100 \times 10$ nm$^3$. The interlayer Co/Pt-Co exchange stiffness was chosen as $A_{Co/Pt-Co} = 2.0 \times 10^{-11}$ J/m.

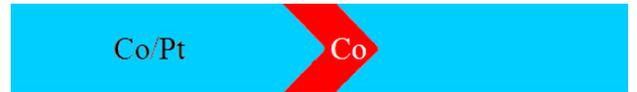

**FIG. 5.** Schematic drawing of V-shaped Co NP on top of Co/Pt NW. The NP is indicated by red color.

The V-shaped NP has several stable magnetic states, which can be switched by external magnetic field. In the first configuration (configuration A) the average NP magnetic moment is directed along the NW (from left to right).

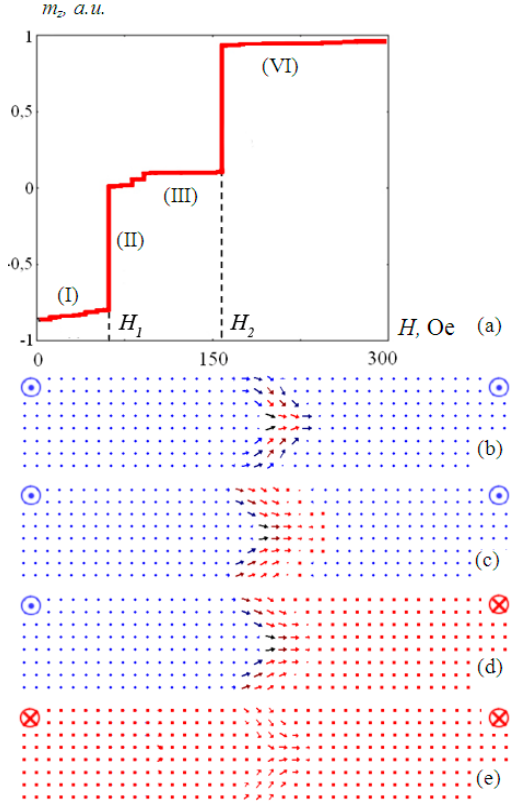

FIG. 6. (a) The $m_z(H)$ curve for A configuration. (b)-(e) The sequential stages of NW magnetization changes under reversed magnetic field corresponding to the transitions I - VI of $m_z(H)$ curve respectively.

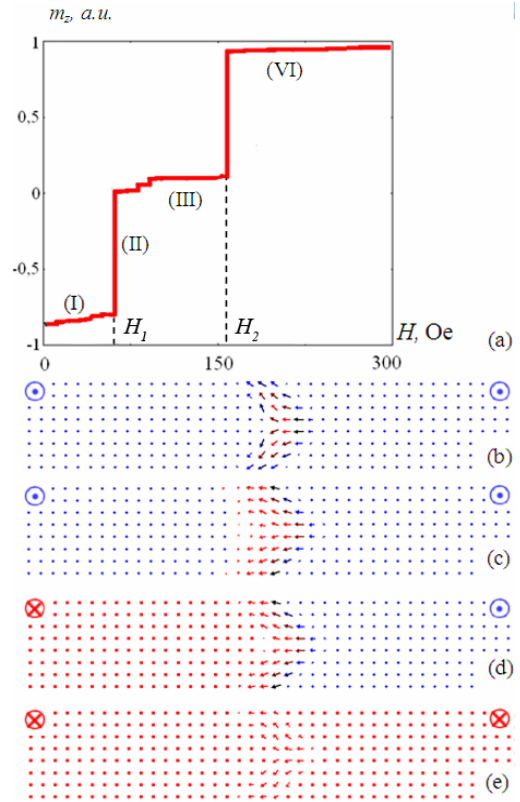

FIG. 7. (a) The $m_z(H)$ curve for B configuration. (b)-(e) The sequential stages of NW magnetization changes under reversed magnetic field corresponding to the transitions I - VI of $m_z(H)$ curve respectively.

The simulated $m_z(H)$ curve and corresponding sequential stages of the NW magnetization reversal for this case are presented in Fig. 6(a-e). The micromagnetic modeling shows that magnetization switching starts at the field $H_1 = 60$ Oe in the area near the NP "tip" (Fig. 6(c)) because of the local NP magnetic field and the remagnetization of NW right part is realized (Fig. 6(d)). Further the system remains stable in external magnetic field up to $H_2 = 160$ Oe and when $H > H_2$ the remagnetization of the rest NW part is observed (Fig. 6 (e)).

If NP magnetic moment has opposite direction (configuration B), as it is shown in the Fig. 7, the NW remagnetization begins in the region near the NP "tail". The critical fields $H_1$ and $H_2$ are the same as in previous case.

The third possible situation is shown in Fig. 8. The average NP magnetic moment is directed across the NW (configuration C). In this case, the NW magnetization reversal starts from the region adjoined to the NP "tail" and at the field $H_3 = 60$ Oe the remagnetization of the left NW part is realized. Further at the field $H_4 = 120$ Oe the NW is remagnetized completely (Fig. 8(e)). Note that the field $H_4$ is appreciably lower than the field $H_2$ for the previous two magnetic configurations.

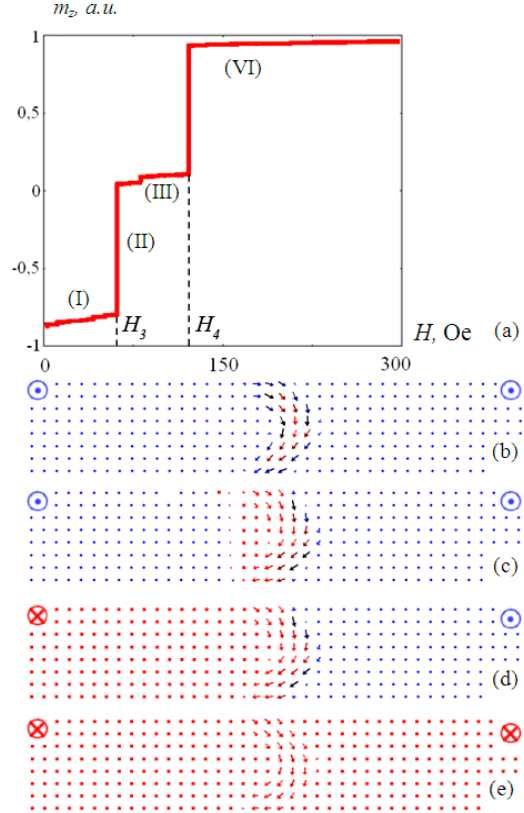

FIG. 8. (a) The $m_z(H)$ curve for C configuration. (b)-(e) The sequential stages of NW magnetization changes under reversed magnetic field corresponding to the transitions I - VI of $m_z(H)$ curve respectively.

## IV. LOGICAL CELLS

Considered above NW-NP system with V-shaped NP can be used for the development of different magnetic LCs. The input information of LC is coded as the direction of NP magnetization. The output information is coded as the magnetization direction at the right NW end. The LC operation is based on the following algorithm. In the first stage the whole system is magnetized uniformly by the strong vertical magnetic field (for definiteness, we assume that the system is magnetized upward). Then the input information is written to the NP using a local in-plane magnetic field. After that the testing magnetic field $H_T$ (perpendicular to the sample plane) with magnitude $H_1 < H_T < H_2$ is applied to the system and output information is read.

TABLE I

| Input NP magnetization | Output NW magnetization | Boolean value |
|---|---|---|
| ↗ | ⊙ | "0" |
| ↘ | ⊗ | "1" |

**Table 1.** The correspondence between directions of input and output magnetization and logical "0" and "1".

In particular, the simplest LC performing operation «NOT» is shown in Fig. 9(a). The corresponding truth table is presented in Fig. 9(b).

The other LCs having more complicated geometry are shown in Fig. 10-12. These LCs have Y-shaped NW and two input NPs. The logical function of these LCs is determined by orientation of NPs relative to NW. As an example we offer some LCs performing operation of "IMPLIFICATION" (Fig. 10), "OR" (Fig. 11) and "NOT-AND" (Fig.12). All these LCs were tested with OOMMF simulator.

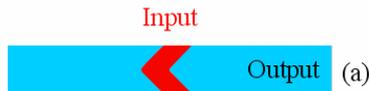
(a)

| Input | Output |
|---|---|
| 1 | 0 |
| 0 | 1 |

(b)

**FIG. 9.** (a) The schematic drawing of LC performing "NOT" operation. (b) Corresponding truth table.

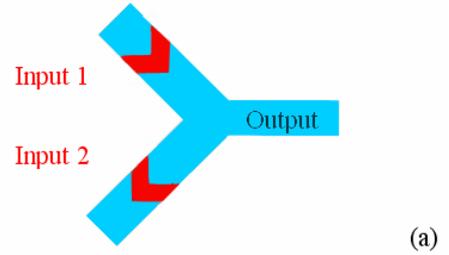
(a)

| Input 1 | Input 2 | Output |
|---|---|---|
| 1 | 1 | 1 |
| 1 | 0 | 1 |
| 0 | 1 | 0 |
| 0 | 0 | 1 |

(b)

**FIG. 10.** (a) LC performing logical operation "IMPLIFICATION". (b) Corresponding truth table.

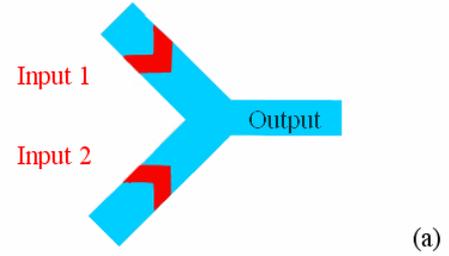
(a)

| Input 1 | Input 2 | Output |
|---|---|---|
| 1 | 1 | 1 |
| 1 | 0 | 0 |
| 0 | 1 | 0 |
| 0 | 0 | 0 |

(b)

**FIG. 11.** (a) LC performing logical operation "OR". (b) Corresponding truth table.

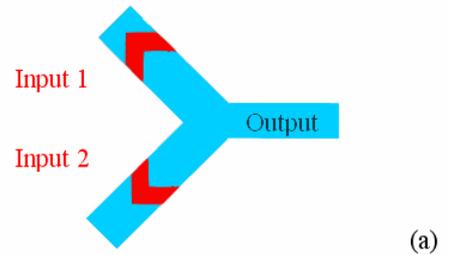
(a)

| Input 1 | Input 2 | Output |
|---|---|---|
| 1 | 1 | 1 |
| 1 | 0 | 1 |
| 0 | 1 | 1 |
| 0 | 0 | 0 |

(b)

**FIG. 12.** (a) LC performing logical operation "AND-NOT". (b) Corresponding truth table.

## V. Conclusion

Thus in present paper we considered the peculiarities of the magnetization reversal in the planar ferromagnetic system consisting of exchange coupled NW with perpendicular magnetization and V-shaped NP with in-plane magnetization. It was shown that the character of NW magnetization reversal and magnitudes of critical fields essentially depends on the direction of NP magnetization. Several schemes of magnetic LCs performing various logic operations were proposed.

This work was supported by the Russian Foundation for Basic Research (Projects No. 15-42-02388 and 15-02-04462).

## References

[1] L. O'Brien, D. E. Read, H. T. Zeng, E. R. Lewis, D. Petit, and R. P. Cowburn, "Bidirectional magnetic nanowire shift register," *Appl. Phys. Lett.*, vol. 95, p. 232502, 2009.

[2] H. T. Zeng, D. E. Read, L. O'Brien, J. Sampaio, E. R. Lewis, D. Petit, and R. P. Cowburn, "The influence of wire width on the charge distribution of transverse domain walls and their stray field interactions," *Appl. Phys. Lett.*, vol. 96, p. 262510, 2010.

[3] R. Mattheis, S. Glathe, M. Diegel, and U. Hubner, "Concepts and steps for the realization of a new domain wall based giant magnetoresistance nanowire device: From the available $2^4$ multiturn counter to a $2^{12}$ turn counter," *J. Appl. Phys.*, vol. 111, p. 113920, 2012.

[4] H. T. Zeng, D. Petit, L. O'Brien, D. Read, E. R. Lewis, R. P. Cowburn, "Tunable remote pinning of domain walls in magnetic nanowires," *J. Magn. Magn. Mater.*, vol. 322, p. 262510, 2010.

[5] L. O'Brien, D. Petit, E. R. Lewis, R. P. Cowburn, D. E. Read, J. Sampaio, H. T. Zeng, and A.-V. Jausovec, "Tunable remote pinning of domain walls in magnetic nanowires," *Phys. Rev. Lett.* vol. 106, p. 087204, 2011.

[6] M. Hayashi, L. Thomas, R. Moriya, C. Rettner, and S. S. P. Parkin, "Current-controlled magnetic domain-wall nanowire shift register," *Science*, vol. 320, p. 209, 2008.

[7] E. R. Lewis, D. Petit, L. Thevenard, A. V. Jausovec, L. O'Brien, D. E. Read, and R. P. Cowburn, "Magnetic domain wall pinning by a curved conduit," *Appl. Phys. Lett.*, vol. 95, p. 152505, 2009.

[8] D. Petit, A. V. Jausovec, D. E. Read, and R. P. Cowburn, "Domain wall pinning and potential landscapes created by constrictions and protrusions in ferromagnetic nanowires," *J. Appl. Phys.*, vol. 103, p. 114307, 2008.

[9] D. Petit, A. V. Jausovec, H. T. Zeng, E. R. Lewis, L. O'Brien, D. E. Read, and R. P. Cowburn, "Mechanism for domain wall pinning and potential landscape modification by artificially patterned traps in ferromagnetic nanowires," *Phys. Rev. B*, vol. 79, p. 214405, 2009.

[10] K. O'Shea, S. McVitie, J. N. Chapman, and J. M. R. Weaver, "Direct observation of changes to domain wall structures in magnetic nanowires of varying width" *Appl. Phys. Lett.*, vol. 93, p. 202505, 2008.

[11] L. K. Bogart, D. Atkinson, K. O'Shea, D. McGrouther, and S. McVitie, "Dependence of domain wall pinning potential landscapes on domain wall chirality and pinning site geometry in planar nanowires," *Phys. Rev. B*, vol. 79, p. 054414, 2009.

[12] E. Varga, G. Csaba, A. Imre, and W. Porod "Simulation of magnetization reversal and domain-wall trapping in submicron permalloy wires with different wire geometries", *IEEE Trans. Nanotechnol.*, vol. 11, p. 682, 2012.

[13] Q. Zhu, X. Liu, S. Zhang, Q. Zheng, J. Wang, and Q. Liu, "Current-induced domain wall motion in nanostrip-nanobars system," *Jpn. J. Appl. Phys.*, vol. 53, p. 073001, 2014.

[14] S. M. Ahn, K. W. Moon, C. G. Cho, and S. B. Choe, "Control of domain wall pinning in ferromagnetic nanowires by magnetic stray fields" *Nanotechnology*, vol. 22, p. 085201, 2011.

[15] L. Sun, R. X. Cao, B. F. Miao, Z. Feng, B. You, D. Wu, W. Zhang, An Hu, and H. F. Ding, "Creating an artificial two-dimensional skyrmion crystal by nanopattern," *Phys. Rev. Lett.*, vol. 110, p. 167201, 2013.

[16] V. L. Mironov, O. L. Ermolaeva, E. V. Skorohodov, and A. Yu. Klimov, "Field-controlled domain wall pinning-depinning effects in a ferromagnetic nanowire-nanoislands system," *Phys. Rev. B*, vol. 85, p. 144418, 2012.

[17] V. L. Mironov, O. L. Ermolaeva, "Domain wall pinning in a ferromagnetic nanowire by stray fields of nanoparticles," *Bulletin of the Russian Academy of Sciences: Physics*, vol. 78, no. 1, p. 16, 2014.

[18] O. L. Ermolaeva, E. V. Skorokhodov, V. L. Mironov, "Domain wall pinning controlled by the magnetic field of four nanoparticles," *Physics of the Solid State,* vol. 58, no. 11, p. 2223, 2016.

[19] V. L. Mironov, O. L. Ermolaeva, E. V. Skorohodov "Controlled domain wall pinning in ferromagnetic nanowire by nanoparticles' stray fields", *IEEE Trans. Magn.*, vol. 52, no. 12, p. 1100607, 2016.

[20] V. L. Mironov, B. G. Gribkov, S. N. Vdovichev, S. A. Gusev, A. A. Fraerman, O. L. Ermolaeva, A. B. Shubin, A. M. Alexeev, P. A. Zhdan, and C. Binns, "Magnetic force microscope tip induced remagnetization of CoPt nanodiscs with perpendicular anisotropy," *J. Appl. Phys.*, vol. 106, p. 053911, 2009.

[21] R. A. van Mourik, C. T. Rettner, B. Koopmans and S. S. P. Parkin, "Control of domain wall pinning by switchable nanomagnet state," *J. Appl. Phys,.* vol. 115, p. 17D503, 2014.

[22] M. J. Donahue and D. G. Porter. OOMMF User's Guide. National Institute of Standards and Technology, Gaithersburg, MD, USA. [Online]. Available: http://math.nist.gov/oommf.

[23] V. Baltz, A. Marty, B. Rodmacq, B. Dieny, "Magnetic domain replication in interacting bilayers with out-of-plane anisotropy: Application to Co/Pt multilayers," *Phys. Rev. B*, vol. 75, p. 014406, 2007.

[24] E. S. Demidov, N. S. Gusev, L. I. Budarin, E. A. Karashtin, V. L. Mironov and A. A. Fraerman "Interlayer interaction in multilayer [Co/Pt]$_n$/Pt/Co structures", *J. Appl. Phys.*, vol. 120, p. 173901, 2016.